\documentclass[letterpaper]{article}

\usepackage[]{aaai2027}
\usepackage[hyphens]{url}
\usepackage{graphicx}
\usepackage{natbib}
\usepackage{caption}
\usepackage{algorithm}
\usepackage{algorithmic}

\usepackage{booktabs}
\usepackage{amsmath}
\usepackage{amssymb}
\usepackage{xcolor}
\usepackage{tcolorbox}
\usepackage{tikz}
\usetikzlibrary{arrows.meta, fit, backgrounds, positioning, shapes.geometric, calc}

\title{OmniEye: Efficient Multimodal Forensic Video Intelligence for Law-Enforcement Body-Worn Cameras}

\author{
Mamadou K. Keita$^{1}$,
Angela Srbinovska$^{1}$,
Anita Srbinovska$^{2}$,
Nishka Desai$^{2}$,
Isabella Zicari$^{3}$,
P. Kwaku Sanaah-Faried$^{1}$,
Sanjay Charitesh Makam$^{1}$,
Wyatt Auten$^{3}$,
Vivek Senthil$^{1}$,
Hannah Desnick$^{3,5}$,
Jonathan Bateman$^{1}$,
Adrian Martin$^{2}$,
Christopher Homan$^{1}$,
John McCluskey$^{3}$,
Ernest Fokoué$^{1}$
}
\affiliations{
$^{1}$Rochester Institute of Technology, 
$^{2}$Rochester (NY) Police Department, 
$^{3}$University at Albany School of Criminal Justice,
${^4}$University at Albany
$^{5}$New York State Youth Justice Institute
}

\begin{document}

\maketitle

\begin{abstract}
We introduce OmniEye, a multimodal video intelligence system for law-enforcement training and review (source code available on request to verified law-enforcement and public-safety agencies). OmniEye ingests body-worn camera footage and perceives every 30-second window jointly across video and audio with one multimodal foundation model. It then stores the model's structured output in an embedded SQLite database with BM25 full-text search. Officers can question the footage through an agent that writes structured queries, retrieves candidate windows, and re-perceives them with the model before it may cite them. The whole system runs on one 16 GB GPU with a 4-bit quantization-aware-trained model, and it also scales to full bf16 precision on a multi-GPU cluster. 
\end{abstract}

\section{Introduction}
\label{sec:intro}
Modern policing produces video at a scale that no review team can watch. One officer's body-worn camera can record several hours per shift, and a mid-size agency can accumulate tens of thousands of recordings per year \cite{camp2024}. Most of this footage is uneventful, but the small fraction that matters, such as uses of force, pursuits, medical emergencies, and misconduct, is the content that oversight, training, and public-records requests depend on. Current practice, however, relies on slow human review or on incomplete keyword searches of written reports, neither of which is efficient at scale.

There were some initial attempts to use AI on this kind of footage \cite{srbinovska2025towards,srbinovska2025beyond,srbinovska2026ontology,srbinovska2026visual}. Another obvious answer is sending the footage to a cloud video-understanding service, but this is not doable here because the footage contains personally identifiable information and is subject to non-disclosure and chain-of-custody obligations \cite{casey2011digital}, and in many jurisdictions it may not be sent to third-party infrastructure at all. A system for this setting must therefore run entirely on agency-controlled hardware, must not assume availability of a data center, and must produce outputs that can resist tampering.

We designed OmniEye within these constraints. Accordingly, it installs on a workstation or a local GPU and runs interactively or as a batch process. We first apply one open-weights multimodal foundation model (Gemma 4 12B) \cite{gemmateam2026gemma4technicalreport} to every window of every recording. We then build everything else: storage, search, dialogue, reporting, and integrity, so that the model's outputs become durable, queryable, and defensible.

OmniEye supports three use cases. \emph{(U1) Unconditional indexing.} At ingest time, with no user query present, we perceive every window of every recording and assign a category, an anomaly score, free-text descriptions, a transcript, and a set of flags. This pass makes the archive searchable, and it also gives archive-level distributions for reporting. \emph{(U2) Query-conditional question answering.} A reviewer asks a question in natural language, and an agent then retrieves candidate windows from the U1 index, re-perceives them with the question in focus, and answers with citations to specific windows. \emph{(U3) Human correction.} Officers correct any field of any window, non-destructively and with attribution. The first use case feeds the second directly, because the fields written at ingest \emph{are} the retrieval index: BM25 \cite{robertson2009probabilistic} ranks the free text while the structured fields supply the filters. The unconditional pass therefore supplies recall and the query-conditional pass precision, since a candidate surfaced by the index is confirmed against the raw pixels and audio before it may be cited.

Achieving all this (from building to using) on local hardware is nonetheless challenging. We process about a thousand image frames plus raw audio per minute of footage, over an archive of tens of thousands of recordings, on hardware ranging from a single 16 GB GPU to a shared two-GPU cluster partition with a hard 64 GB host-memory cap. The model must therefore fit, must run fast enough to clear a backlog, and must survive multi-hour, thousand-file runs on preemptible, memory-capped nodes without losing work. A large part of our contribution is consequently the set of techniques that make this feasible (Section~\ref{sec:engine}).

Our contributions are the following:
\textbf{(1)} We build a resource-friendly, localizable perception engine (Section~\ref{sec:engine}). It combines 4-bit quantization-aware-trained weights \cite{jacob2018quantization}, per-GPU data-parallel replicas, a lossless temporal speculative decoder that leverages the token overlap between consecutive windows \cite{saxena2023prompt,yang2023llma}, and an additive-increase, multiplicative-decrease (AIMD) batch controller \cite{chiu1989analysis} that adapts to both GPU out-of-memory events and host-memory pressure.
\textbf{(2)} We then build an integrity architecture (Section~\ref{sec:integrity}): every window is bound into a per-video SHA-256 hash chain \cite{nist2015sha,haber1991how}, every mutating action is written to a global append-only, hash-linked, attributed audit log, and every human correction is non-destructive and replayable.
\textbf{(3)} On top of this, we build a re-perception-grounded question-answering agent (Section~\ref{sec:agent}). The citation mechanism is enforced structurally, so the agent may cite a window only after it has re-run the model on that window's pixels and audio in the current turn. This architecture and this grounding rule together follow the notions of \emph{experiment} and \emph{agent non-repudiation} of Keita and Homan \cite{keita2026nonrepudiation}, so that the machine's findings, the humans' corrections, and the agent's claims all sit in one tamper-evident, attributed record.
\textbf{(4)} Finally, we provide a flexible, robust operational framework (Section~\ref{sec:store} and Appendix~\ref{app:ops}).

Finally, we describe an experiment conducted in collaboration with the Rochester Police Department (Section~\ref{sec:experiment}). It evaluates the system's categories and its ranking against human annotators on a year-stratified sample of 1,000 recordings drawn from a population above 90,000.

\section{OmniEye}
\label{sec:omnieye}
Four principles shape every component in OmniEye. \emph{(P1) Local, no cloud.} All computation and storage stay local, and the only network use is optional retrieval of a source URL that a user provides. \emph{(P2) One workspace is one file.} A workspace is a single SQLite database. It holds either one incident, which typically covers several recordings from several officers and cameras, or an entire archive batch. Since it is one file, it is therefore easy to copy, archive, and air-gap. \emph{(P3) Every claim is grounded and auditable.} Nothing the system asserts about footage exists without a window identifier, a timestamp, and a tamper-evident hash lineage. \emph{(P4) Degrade, never fail.} On constrained or preempted hardware the system consequently slows down and resumes rather than crashing or losing work.

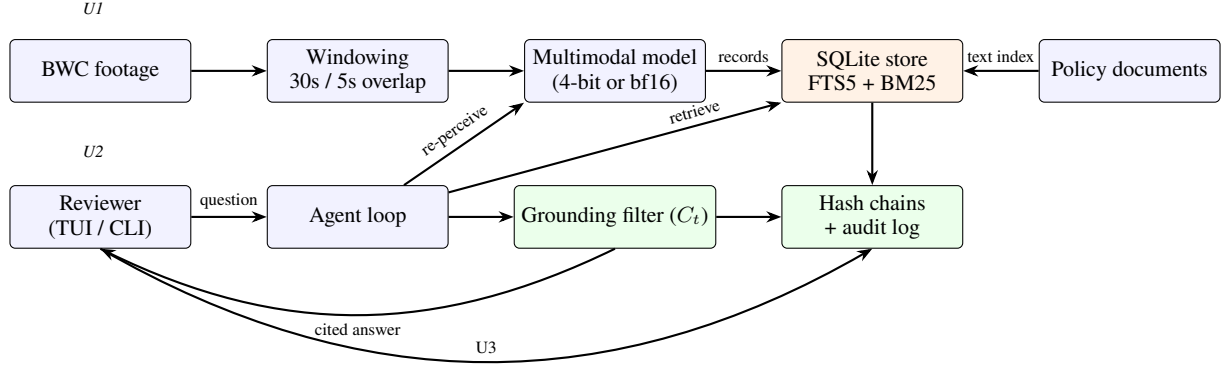
\begin{figure*}[t]
\centering
\small
\begin{tikzpicture}[
  scale=0.92, every node/.style={transform shape},
  box/.style={draw, rounded corners=2pt, align=center, minimum height=9mm, minimum width=26mm, font=\small, fill=blue!5},
  store/.style={draw, rounded corners=2pt, align=center, minimum height=9mm, minimum width=26mm, font=\small, fill=orange!10},
  sec/.style={draw, rounded corners=2pt, align=center, minimum height=9mm, minimum width=26mm, font=\small, fill=green!8},
  lbl/.style={font=\scriptsize\itshape},
  arr/.style={-{Stealth[length=2mm]}, thick}
]

\node[lbl] at (-0.1, 3.0) {U1};
\node[box]   (video) at ( 0.0, 2.1) {BWC footage};
\node[box]   (win)   at ( 3.7, 2.1) {Windowing\\30s / 5s overlap};
\node[box]   (model) at ( 7.4, 2.1) {Multimodal model\\(4-bit or bf16)};
\node[store] (db)    at (11.1, 2.1) {SQLite store\\FTS5 + BM25};
\node[box]   (docs)  at (14.8, 2.1) {Policy documents};

\node[lbl] at (-0.1, 0.95) {U2};
\node[box] (user)   at ( 0.0, 0.0) {Reviewer\\(TUI / CLI)};
\node[box] (agent)  at ( 3.7, 0.0) {Agent loop};
\node[sec] (ground) at ( 7.4, 0.0) {Grounding filter ($C_t$)};
\node[sec] (chain)  at (11.1, 0.0) {Hash chains\\+ audit log};

\draw[arr] (video) -- (win);
\draw[arr] (win) -- (model);
\draw[arr] (model) -- node[above,font=\scriptsize]{records} (db);
\draw[arr] (docs) -- node[above,font=\scriptsize]{text index} (db);
\draw[arr] (db) -- (chain);
\draw[arr] (user) -- node[above,font=\scriptsize]{question} (agent);
\draw[arr] (agent) -- node[above,font=\scriptsize,pos=0.75,sloped]{retrieve} (db.south west);
\draw[arr] (agent) -- node[above,font=\scriptsize,sloped]{re-perceive} (model.south west);
\draw[arr] (agent) -- (ground);
\draw[arr] (ground) -- (chain);
\draw[arr] (ground.south) to[bend left=26] node[below,font=\scriptsize,pos=0.5]{cited answer} (user.south);
\draw[arr] (user.south) to[bend right=30] node[above,font=\scriptsize,pos=0.5]{U3} (chain.south);
\end{tikzpicture}
\caption{OmniEye architecture. The top row is the ingest path (U1) and the bottom row is the query path (U2).}
\label{fig:arch}
\end{figure*}

Figure~\ref{fig:arch} shows the system diagram. On the ingest path (U1), recordings enter the perception pipeline (Section~\ref{sec:pipeline}). It windows them, applies the multimodal perception model to the frames and audio of each window, and then commits one structured record per window into the embedded store (Section~\ref{sec:store}). Ingested policy documents also reach the store on this path, but we chunk and index them as text only. The store in turn indexes free text for BM25 retrieval and binds every record into per-video hash chains (Section~\ref{sec:integrity}). On the query path (U2), the agent (Section~\ref{sec:agent}) retrieves candidates from that index, re-perceives them with the same model and the reviewer's question in focus, and then answers with citations that a grounding filter admits only for re-perceived windows. On the correction path (U3), a reviewer edits fields directly. In every case, whether human or agent, each change appends to a global attributed audit log.

\subsection{Perception Pipeline}
\label{sec:pipeline}

\textbf{Windowing} \\
We segment each recording into fixed windows of length $w = 30$ seconds with an overlap of $o = 5$ seconds, which gives a stride of 25 seconds (Appendix~\ref{app:windowing}). The overlap prevents an event that spans a boundary from being split across two windows without context. We then give each window a deterministic index and identifier (\texttt{\{video\_id\}\_chunk\_\{index\}}), which makes ingestion idempotent, so re-running a video recomputes only the windows whose records are missing. From each window we finally sample frames at 1 frame/s (30 frames) and extract the corresponding 16 kHz mono audio.

\textbf{Joint perception} \\
We pass the frames and raw audio of a window to the model together. A forensic-analyst system prompt then enforces a what/where/when/how grounding logic, so the model asserts only what it can see or hear, marks anything else not determinable, and returns one strict JSON object. We analyze vision and audio jointly, in one forward pass, because this lets the model reason about cross-modal relations that a pipeline of separate detectors cannot, for example that calm imagery accompanies shouted audio. We accordingly keep this as a \texttt{cross\_modal\_tension} signal. Each window then produces one record holding free-text visual, audio, and cross-modal descriptions, a speaker-tagged transcript, a category label with a three-way score distribution, keyword tags, audio events, evidence flags, structured detections, and an \texttt{anomaly\_score} in $[0,1]$ (Appendix~\ref{app:schema}). We also make parsing tolerant: a single-pass repair closes truncated JSON and forces every field, so a malformed generation still outputs its completed fields and we drop no window.\footnote{This is an engineering choice based on observed failures in early implementations.}

\textbf{Classification} \\
We classify every window into one of three categories, which we define to the model verbatim in the prompt: \emph{Boring} (routine patrol, static scenes, no significant interaction), \emph{Could be interesting} (moderate activity such as interactions, minor incidents, and non-critical force or de-escalation), and \emph{Must absolutely watch} (critical event such as significant use of force, arrests, weapons drawn, misconduct, medical emergencies, pursuit, or high legal and public interest). We ask for a score vector $s \in \Delta^2$ rather than only a hard label, so that reviewers can threshold by confidence. Afterwards, we renormalize the scores to sum to one, and we snap a label outside the allowed set to the nearest valid category.

\textbf{Locating events of interest} \\
We do not leave the job of finding the moments that matter to the category alone. Instead, we attach five redundant signals to each window: the category and its score distribution, the \texttt{anomaly\_score}, open-vocabulary evidence flags and audio events, structured booleans, and \texttt{cross\_modal\_tension}. A reviewer or the agent can therefore locate an event by category, by anomaly ranking, by a structured flag, by a free-text cue, or by cross-modal conflict, and more than one path usually catches the same event (Appendix~\ref{app:signals}).

\textbf{Summarization} \\
After all windows of a video are perceived, we then build a two-level narrative over segments of ten windows each. Appendix~\ref{app:summ} describes it.

\subsection{Perception Engine}
\label{sec:engine}

We run the full multimodal model locally, not behind an inference server. For a unified audio and vision architecture on our target hardware, external serving frameworks either lacked correct model support or added overhead we could not afford. Local execution also lets ingestion and interactive chat share one copy of the weights. On top of this, we apply four techniques, and each one targets one constraint (Appendix~\ref{app:eff}).

\textbf{(1) Memory resources} \\
To fit a 12B multimodal model\footnote{We use Gemma 4 12B \cite{gemmateam2026gemma4technicalreport} to power OmniEye.} plus a 30-frame-and-audio activation peak inside a 16 GB consumer GPU, we load the model with 4-bit NF4 weight quantization and double quantization \cite{dettmers2023qlora,dettmers2022llmint8}. We nonetheless keep the multimodal embedder and the output head in higher precision. We also quantize a quantization-aware-trained INT4 checkpoint \cite{jacob2018quantization}, so 4-bit inference tracks the model's trained operating point rather than degrading a bf16 model after the fact. Together, this brings the footprint to roughly 6 to 8 GB and preserves audio and vision accuracy, which post-training INT4 of the embedders does not. On cluster GPUs we instead load full bf16.

\textbf{(2) Throughput per GPU} \\
Perception is decode-bound, since 4-bit dequantize-matmul sustains only about 9 tokens/s of greedy decode in our measurements. Consecutive windows, however, describe overlapping scenes, so they share a large fraction of their output tokens. We therefore leverage this with a prompt-lookup speculative decoder \cite{saxena2023prompt,yang2023llma,leviathan2023fast,chen2023accelerating}, and we use the sequence emitted for window $i{-}1$ in the same stream as the draft reference. Since the model verifies every proposed token, \textbf{the output is byte-identical to greedy decoding}, so the speedup carries no quality change. In practice we measured $\bar{\alpha} \approx 2.4$ accepted tokens per verification forward and a $1.48\times$ window speedup. The scheme moreover needs no draft model, no extra weights, and no additional memory (Appendices~\ref{app:enginedetail} and~\ref{app:spec}).

\textbf{(3) Throughput across GPUs} \\
Batching across windows shows no gain for 4-bit inference, since a batch of two measured $1.03\times$, so the effective multi-GPU lever is instead data parallelism. We load one independent model replica per visible GPU and partition a video's windows into one contiguous stream per replica, so $N$ GPUs give about $N\times$ ingest throughput. Adjacency matters here because the speculative decoder needs each replica to see consecutive windows. In addition, a scheduler lets interactive chat preempt ingestion between windows, which drops chat latency during ingest from a whole window to roughly one forward pass while ingest still completes losslessly (Appendix~\ref{app:enginedetail}).

\textbf{(4) Cluster throughput} \\
On data-center GPUs such as the H100, the constraint changes. There we load full bf16 and we do batch multiple windows per forward, because at bf16 the batched matmul is bandwidth-bound rather than launch-bound. We cannot, however, choose a batch size ahead of time on multiprogrammed GPUs, where the partition is shared with other jobs and the free memory available to us varies at run time, so we control it instead with an AIMD loop \cite{chiu1989analysis}. The controller halves the batch on a CUDA out-of-memory error, retries the same windows, and then grows back after three clean batches (Appendix~\ref{app:enginedetail}). We lose nothing on such an error, so a tight or shared GPU recovers automatically.

\textbf{Host memory} \\
On shared clusters the constraint is often not GPU memory but host memory, and a thousand-video run is killed after a few dozen videos. We therefore apply the same AIMD idea to host RAM, so a 1,000-video bf16 run self-paces within a fixed 64 GB cap with no work lost. Under pressure it consequently processes smaller batches, never fewer windows (Appendix~\ref{app:enginedetail}).

\subsection{Storage and Search}
\label{sec:store}
We keep all state in one SQLite database \cite{sqlite} with write-ahead logging, with tables for videos, segments, windows, ingested documents, corrections, the audit log, and reports. We make the file the workspace boundary, so a case is a single portable artifact, and because perception commits each window as it completes, the database is always a valid checkpoint. Beyond that, a workspace exports to a single \texttt{.omnieye} archive (tar plus Zstandard \cite{rfc8878}) that unpacks with no reprocessing and records unfinished jobs, so a case can move between machines and continue ingesting (Appendix~\ref{app:ops}).

For search, we index the free-text fields of every window (\texttt{description}, \texttt{audio\_description}, \texttt{cross\_modal}, \texttt{transcript}) in an FTS5 virtual table with Porter stemming \cite{porter1980algorithm}, and we then rank by Okapi BM25 \cite{robertson2009probabilistic} (Appendix~\ref{app:bm25}). We further combine this lexical ranking with structured filters (category, tags, evidence flags and audio events, boolean detections, officer, environment, anomaly threshold, and time range), so a query such as ``weapon-drawn windows after 12 minutes with an anomaly above 0.6'' is one SQL statement. Finally, we chunk ingested policy documents to paragraph granularity and index them the same way, which makes department policy citable alongside footage.

\subsection{Integrity and Non-Repudiation}
\label{sec:integrity}
OmniEye's outputs may end up in an important task, so we treat integrity as a must have feature.

\textbf{Hash chains} \\
We bind each window record into a per-video chain in the style of linked timestamping \cite{haber1991how,merkle1987digital}. Specifically, for window $i$ we compute
\[
h_i \;=\; \mathrm{SHA256}\big(\,\mathrm{canon}(\,f,\; i,\; o_i,\; t_i,\; h_{i-1}\,)\,\big),
\]
where $f$ is the source file's hash, $o_i$ is the raw model output for the window, $t_i$ is the ingest timestamp, $h_{i-1}$ is the previous window's hash (a fixed genesis value $h_{-1} = g$ for $i = 0$), and $\mathrm{canon}(\cdot)$ is a canonical, key-sorted JSON serialization \cite{nist2015sha}. Each record then stores both $h_{i-1}$ and $h_i$, and verification checks in order that each recorded predecessor hash equals the actual predecessor's hash, which in turn detects reordering, insertion, deletion, or modification of any window and localizes the break to a window index (Figure~\ref{fig:chain}).

\textbf{Audit log} \\
We also append every mutating action, most importantly every human correction, to a second global hash-linked log. Entry $j$ stores the hash of the prior entry and a content hash $c_j$ over its own canonical payload, which covers the acting officer's identity and a timestamp. Verification then recomputes both and flags a broken chain or a content mismatch by sequence number (Appendix~\ref{app:nonrep}). Together, the two chains make both the machine's findings and the humans' edits tamper-evident.

\textbf{Non-repudiation} \\
Tamper-evidence alone, however, is not enough for an evidentiary setting. Non-repudiation means an actor cannot later deny an action they took \cite{iso27000}. We therefore follow Keita and Homan \cite{keita2026nonrepudiation} and implement it by binding the identity of the actor, the exact content of the action, and its position in the append-only chain into one record that no one can quietly change (Appendix~\ref{app:nonrep}).

\textbf{Corrections} \\
Officers can correct any field of any window or segment, but we never overwrite the original model output. Instead, we store corrections as append-only records with an action, the old and new values, a note, and the reviewer's identity, and each also writes an audit entry. We then reconstruct the effective view of a window by replaying its corrections over the stored original, so the full provenance is always recoverable (Appendix~\ref{app:nonrep}).

\subsection{Re-Perception-Grounded Agent}
\label{sec:agent}

We answer natural-language questions with a tool-calling loop \cite{yao2023react,schick2023toolformer} over the store. The architecture is that of retrieval-augmented generation \cite{lewis2020rag}: the agent retrieves candidate windows and then composes a textual answer whose claims carry citations, each citation being a window identifier and its timestamp, rendered in the terminal application as a clickable, playable reference to that stretch of footage.

\textbf{Re-perception} \\
Search returns candidate windows, but a candidate is not yet citable. The agent may cite a window only after it has re-perceived it in the current turn, meaning it has re-run the model on that window's frames and audio with the question in focus. This serves two purposes. First, the record written at ingest was produced with no question present, so it describes the window generically and may not mention what a later question asks about. More importantly, re-perception makes an unsupported citation impossible rather than merely discouraged. Formally, let $R_t$ be the set of windows re-perceived during turn $t$ and $\mathrm{cites}(y_t)$ the citation markers in the drafted answer $y_t$. The system then produces the filtered answer
\[
y_t' \;=\; \mathrm{strip}\big(\, y_t,\; \mathrm{cites}(y_t) \setminus R_t \,\big),
\]
where $\mathrm{strip}$ deletes from the drafted text every citation marker naming a window outside $R_t$, together with any claim left unsupported once that marker is gone, so the set of valid citations is exactly $C_t = \mathrm{cites}(y_t) \cap R_t$. Since large language models otherwise hallucinate confidently \cite{ji2023survey}, we consider this our main defense in an evidentiary setting. In addition, before an answer is finalized, a grounding verifier also re-reads the drafted text against the records of the windows in $R_t$ under the what/where/when/how rule of Section~\ref{sec:pipeline}, keeping only the claims the evidence supports. As is standard for tool-calling agents, the re-perception loop is moreover bounded: it runs for at most 8 steps and performs at most 6 re-perceptions per question, while read-only retrieval and analysis calls are unbudgeted. Appendix~\ref{app:detectors} passes one question through the loop.

\textbf{Agent non-repudiation} \\
We implement agent non-repudiation with the audit log of Section~\ref{sec:integrity} as the record, and with re-perceptions, answers, and citations as the designated actions. Concretely, we append each one with the agent as the attributed actor, carrying the question, the window identifiers in $R_t$, and content hashes over the re-perception outputs and the filtered answer $y_t'$. Since each entry is chained through $\ell_j$, the three denials that agent non-repudiation rules out would each therefore require a detectable chain break (Appendix~\ref{app:integrity}). As a result, the machine's findings, the humans' corrections, and the agent's claims live in one tamper-evident, attributed structure.

\section{Evaluation and Discussion}
\label{sec:experiment}

\begin{tcolorbox}[colback=gray!5!white,colframe=gray!75!black,boxsep=1pt,left=3pt,right=3pt,top=2pt,bottom=2pt,title=\textbf{Data disclosure statement}]
\small This project was conducted in collaboration with the Rochester (NY) Police Department under strict guidelines on what may be disclosed. We therefore give no details on the data beyond the sampling procedure and the results reported here.
\end{tcolorbox}

We evaluate whether OmniEye's three categories agree with human annotators, and whether the system orders footage usefully for review.

\textbf{Sampling} \\
Our total population exceeds 90,000 recordings. We cannot process and human-annotate that volume within our constraints, so we instead draw a probability sample of $n = 1{,}000$ stratified by recording year, allocated in proportion to each year's share (Appendix~\ref{app:windowing}).

\textbf{Processing} \\
We ran the complete pipeline over the 1,000 sampled recordings on a node with two H100 GPUs in bf16 cluster mode with adaptive batching (Section~\ref{sec:engine}), under the fixed 64 GB host-memory allocation. Seven of the sampled files, however, had zero duration and showed no windows. We logged and skipped them without interrupting the run, which left 993 recordings processed. The loss occurs after sampling, so we do not reweight. In total, the run produced 17,096 windows, classified 67.4\% \emph{Boring}, 29.1\% \emph{Could be interesting}, and 3.5\% \emph{Must absolutely watch}.

\textbf{Protocol} \\
Within the 1,000 recordings, we randomly sample 500 windows, each annotated by two trained annotators, A and B.\footnote{We annotate windows rather than whole videos because video length varies from minutes to hours, and an annotator who loses focus over that much footage annotates trivially, which would corrupt the comparison.} All 500 fall inside the 993 processed recordings, so annotation and processing saw the same thing, and neither annotator saw the system's output, so every comparison below is blind. For each window an annotator gives a review priority on four levels, \textsc{no\_review\_needed}, \textsc{review\_worthy}, \textsc{high\_priority\_review}, and \textsc{unable\_to\_assess}, with one or more reason categories and a short description of why. Note that \textsc{unable\_to\_assess} has no counterpart in the model's label set, since an obstructed or dark window admits no priority judgment. We then map \textsc{no\_review\_needed} to \emph{Boring}, \textsc{review\_worthy} to \emph{Could be interesting}, and \textsc{high\_priority\_review} to \emph{Must absolutely watch}. For the binary numbers we further call a window \emph{worth reviewing} if an annotator placed it above \emph{Boring}, and we count \textsc{unable\_to\_assess} as not worth reviewing. Appendix~\ref{app:matrices} shows that the alternatives do not drive the results.

\subsection{Results and Discussion}

\begin{table}[t]
\centering
\footnotesize
\begin{tabular}{@{}lrcc@{}}
\toprule
OmniEye category & $n$ & Annotator A & Annotator B \\
\midrule
Boring                & 332 & 0.44 [0.39, 0.50] & 0.08 [0.05, 0.11] \\
Could be interesting  & 153 & 0.76 [0.68, 0.82] & 0.16 [0.11, 0.23] \\
Must absolutely watch &  15 & 1.00 [0.80, 1.00] & 0.60 [0.36, 0.80] \\
\bottomrule
\end{tabular}
\caption{Windows each annotator marked worth reviewing, by OmniEye category. $n$ is the number of annotated windows in that category, and the confidence interval is 95\% .}
\label{tab:results}
\end{table}

\begin{table}[t]
\centering
\footnotesize
\begin{tabular}{@{}lcccc@{}}
\toprule
Target & Base & P@25 & P@50 & P@100 \\
\midrule
Worth reviewing (A)  & 0.556 & 0.88 & 0.82 & 0.84 \\
Worth reviewing (B)  & 0.118 & 0.48 & 0.38 & 0.29 \\
High priority (A)    & 0.180 & 0.60 & 0.48 & 0.43 \\
\bottomrule
\end{tabular}
\caption{Precision at $k$ for the 500 annotated windows ranked by \texttt{anomaly\_score}.}
\label{tab:patk}
\end{table}

The two annotators worked from the same tool\footnote{The tool is available upon request \cite{OpenBWC_visual_timelines}.} on the same windows, and yet they labelled the sample differently. Exact four-category agreement is 43.0\% and binary agreement is 55.0\%, which gives Cohen's $\kappa = 0.107$ and $\kappa = 0.171$. Annotator A marked 278 of 500 windows worth reviewing and annotator B marked 59, a factor of 4.7. Crucially, we notice the disagreement is in one direction. Of B's 59 positives, 56 are also A's, and only 3 go the other way (Table~\ref{tab:ab}). The two annotators are therefore not reading the footage incompatibly. Rather, they apply different thresholds to what is largely a shared ordering, and B used \textsc{high\_priority\_review} once in 500 windows, so the four-category $\kappa$ partly measures a level one annotator did not use.

As with any heavy class imbalance, $\kappa$ is lowered by default when the marginals are this skewed. For instance, a system that placed every window in \emph{Boring} would match annotator B on 88.2\% of the sample and still be useless. Agreement with either annotator therefore measures closeness to that annotator, not correctness. Against that baseline, OmniEye nonetheless agrees with A more than B does ($\kappa = 0.290$ versus $0.171$). We resample videos rather than windows in a 2,000-sample bootstrap, since the 500 windows come from 338 recordings, and we obtain a difference of $0.119$ with a 95\% interval of $[0.030, 0.200]$.

The agreement is low and one-directional, and the tool's main use case is retrieval rather than labeling, so we ask instead whether the system reaches the footage a reviewer wants quicker. Table~\ref{tab:patk} details this. When we rank the annotated windows by \texttt{anomaly\_score}, the top 25 hold high-priority content at 0.60 against a base rate of 0.180, a $3.3\times$ lift independent of either threshold. On the ROC-AUC, the score similarly reaches $0.677$ against A and $0.686$ against B, ahead of the category used as an ordinal score ($0.657$ and $0.655$), so we read the category as a rough ordering device rather than the system's main signal. Put concretely, at annotator B's threshold, a reviewer going in random order hits one window of interest every 8.5 windows, one every 3.4 in the top fifth by anomaly score, and one every 1.7 inside \emph{Must absolutely watch}. At A's threshold the figures are instead 1.8, 1.2, and 1.0.

Moreover, the ordering persists through the disagreement. In Table~\ref{tab:results} the \emph{Boring} and \emph{Could be interesting} intervals do not overlap for either annotator, so two people whose cutoffs differ by a factor of 4.7 still put the three categories in the same order. The category is nevertheless still unsafe as a filter. A reviewer who skipped \emph{Boring} would lose 147 of A's 278 worth-reviewing windows, including 42 of her 90 \textsc{high\_priority\_review} ones. For that reason, we use the category only to order work, and the agent confirms candidates by re-perception rather than by category.

The disagreement is also concentrated, running from 9.5\% on \textsc{no\_notable\_event} to 62.5\% on \textsc{officer\_subject\_interaction} (Appendix~\ref{app:disagreement}). We held an informal session with departmental reviewers on two further recordings, which suggests why. Reviewers thinking about tactics and reviewers thinking about procedure grouped the same recordings differently, and both readings were right. Whether a window is worth reviewing therefore appears to be a property of the footage plus the reviewer's perspective, which would explain the low $\kappa$ and why the ordering holds across annotators while the labels do not. On those same two recordings we also tested the agent path against an eight-question departmental review form. All sixteen answers carried window-level citations with timestamps, which is the filter $C_t$ of Section~\ref{sec:agent} working end to end (Appendix~\ref{app:detectors}).

\section{Conclusion}
\label{sec:conclusion}

We introduced OmniEye, a tool that applies one open-weights multimodal model to every window of every recording. Around it, we build the storage, the search, the dialogue, and the integrity, so that the model's outputs become durable, queryable, and defensible. It runs entirely on agency hardware, from a single 16 GB consumer GPU to a shared cluster partition under a fixed 64 GB host-memory cap. In addition, every window record sits in a per-video hash chain, every correction is attributed and replayable, and the agent may cite a window only after it has re-perceived it in the same turn.

We evaluated the system on 1,000 recordings from the Rochester Police Department. Two annotators labelled the same 500 windows blind and agreed at $\kappa = 0.171$, and the system agreed with one of them more than the other annotator did. Their disagreement, however, ran almost entirely in one direction, so they shared an ordering and differed on where to cut it. There is thus no single ground truth here, since whether a window is worth reviewing depends on the reviewer's perspective. Evaluated as a ranking instead, the system reaches high-priority content at 3.3 times the base rate in its top 25 windows.

For deployment, OmniEye is not yet in daily production use, and we are rolling it out into the Rochester Police Department in stages. First, we process the archived recordings on department infrastructure, with reviewers annotating the output. Second, we train key staff members to use the tool to identify training gaps. Third, we wire the tool into the department's existing analysis workflow. Last, we make the tool available for any other law enforcement agencies.

\appendix

\section{Acknowledgments}
This research is supported by Grant 15PBJA-22-GG-03328-BWCx with the U.S. Department of Justice through its Office of Justice Programs and Bureau of Justice Assistance, awarded to the City of Rochester, with ongoing support from the College of Science at Rochester Institute of Technology.

Special thanks to Angeliki Perrikos, Itzel Gutierrez Tapia, and Hipolita Oliva for their assistance with reviewing and annotating portions of the data.

\section{Limitations}
\label{app:limitations}

(1) The annotation study uses only two annotators, because the department's guidelines restrict who may access the footage and take part in the project, and two annotators were the most we could obtain. With $\kappa = 0.171$ between them the pattern we report is factual, but we do not know how it looks with a larger pool. The \emph{Must absolutely watch} category likewise rests on 15 windows, and a stratified design with equal allocation across categories, reweighted to population rates, would give a sharper estimate.

(2) The expert session covers two recordings with a few participants and no formal protocol, so it explains rather than measures.

(3) The lack of multiple baselines follows from the same restrictions, both on which models are permitted and on the compute we could afford. The footage also carries personally identifiable information and is under chain-of-custody rules, so it cannot leave agency infrastructure without approval from the jurisdiction, which we did not have and did not seek.

\section{Annotator Disagreement}
\label{app:disagreement}

\begin{table}[h]
\centering
\small
\begin{tabular}{@{}lrc@{}}
\toprule
Reason category & mentions & Disagreement \\
\midrule
\textsc{search\_entry\_or\_room\_clearing} &  28 & 0.68 \\
\textsc{officer\_subject\_interaction}     & 256 & 0.63 \\
\textsc{restraint\_or\_handcuffing}        &  25 & 0.60 \\
\textsc{person\_in\_distress}              &  32 & 0.56 \\
\textsc{medical\_event}                    &  42 & 0.48 \\
\textsc{off\_target\_view}                 &  32 & 0.47 \\
\textsc{routine\_activity}                 & 327 & 0.47 \\
\textsc{normal\_movement}                  & 252 & 0.37 \\
\textsc{low\_visibility}                   &  67 & 0.33 \\
\textsc{waiting\_or\_idle}                 & 176 & 0.31 \\
\textsc{no\_notable\_event}                & 169 & 0.10 \\
\bottomrule
\end{tabular}
\caption{Binary disagreement between the two annotators, by reason category. We count annotator-window mentions summed over both annotators rather than windows, so a window both annotators tagged counts twice. We list the eleven categories with at least 25 mentions. Overall binary disagreement is 45.0\% and four-category disagreement is 57.0\%.}
\label{tab:disagreement}
\end{table}

Beyond this, both annotators also needed a category the three-way label set does not provide, assigning \textsc{unable\_to\_assess} to 34 and 23 windows respectively, for obstructed cameras, off-target views, and severe low light.

In the informal session referred to in Section~\ref{sec:experiment}, \textbf{\textit{reviewers thinking about tactics said the two recordings had nothing in common, while reviewers thinking about procedure and liability grouped them as traffic stops with uncooperative drivers who first refused to get out of the car, with the same officer on both jobs}}. Both interpretations are right, and they differ only in the perspective each reviewer brings.

\section{Category Agreement}
\label{app:matrices}
The binary mapping counts \textsc{unable\_to\_assess} as not worth reviewing, giving $\kappa = 0.171$; dropping those 57 windows instead gives $\kappa = 0.174$ on 454 windows, while counting them as worth reviewing gives $\kappa = 0.191$.

\begin{table}[h]
\centering
\small
\begin{tabular}{@{}lrrrr@{}}
\toprule
& \multicolumn{4}{c}{Annotator B} \\
\cmidrule(l){2-5}
Annotator A & None & Worthy & High & Unable \\
\midrule
\textsc{no\_review\_needed}     & 185 &  1 & 0 &  2 \\
\textsc{review\_worthy}         & 165 & 18 & 0 &  5 \\
\textsc{high\_priority\_review} &  47 & 37 & 1 &  5 \\
\textsc{unable\_to\_assess}     &  21 &  2 & 0 & 11 \\
\bottomrule
\end{tabular}
\caption{The two annotators against each other on all 500 windows.}
\label{tab:ab}
\end{table}

\begin{table}[h]
\centering
\small
\begin{tabular}{@{}lrccc@{}}
\toprule
OmniEye category & $n$ & Boring & Could be int. & Must watch \\
\midrule
\multicolumn{5}{@{}l}{\emph{Annotator A}}\\
Boring                & 305 & 0.52 & 0.34 & 0.14 \\
Could be interesting  & 146 & 0.21 & 0.55 & 0.24 \\
Must absolutely watch &  15 & 0.00 & 0.13 & 0.87 \\
\midrule
\multicolumn{5}{@{}l}{\emph{Annotator B}}\\
Boring                & 314 & 0.92 & 0.08 & 0.00 \\
Could be interesting  & 148 & 0.83 & 0.17 & 0.00 \\
Must absolutely watch &  15 & 0.40 & 0.53 & 0.07 \\
\bottomrule
\end{tabular}
\caption{Annotator categories against OmniEye categories. Rows sum to one, and $n$ gives the row count. We leave out \textsc{unable\_to\_assess} windows, so row totals differ between annotators.}
\label{tab:matrices}
\end{table}

\section{Windowing and Sampling}
\label{app:windowing}
Stratifying a population of size $N_p$ by recording year, stratum $y$ with population $N_y$ receives $n_y = n \cdot N_y / N_p$ of the total sample $n$. Recording years, however, come from a separate metadata table rather than from the workspace, so the year allocation cannot be verified from the exported workspace alone.

Similarly, a recording of duration $T$ has
\[
K \;=\; \Big\lceil \frac{T - o}{w - o} \Big\rceil, \qquad w = 30,\; o = 5,
\]
windows, where window $i \in \{0, \dots, K{-}1\}$ covers the interval $[\, i(w - o),\; \min(i(w-o) + w,\, T) \,)$.

\section{Detection Signals}
\label{app:signals}
The \texttt{cluster} field and its score distribution give a category. The \texttt{anomaly\_score} $a_i \in [0,1]$ then captures unusualness even when the category is conservative. It drives sorting and spike detection, and we accordingly set a per-video flag
\[
\texttt{has\_critical\_spike} \;=\; \mathbb{1}\!\left[\, \max_{i} a_i > 0.70 \,\right].
\]
Evidence flags and audio events are in turn open-vocabulary strings (for example \texttt{hands\_not\_visible}, \texttt{proximity\_violation}, \texttt{gunshot}, \texttt{radio\_static}) that mark forensically relevant cues. Structured booleans (\texttt{persons\_count}, \texttt{has\_weapon}, \texttt{weapon\_drawn}, \texttt{physical\_contact}) give high-precision, directly filterable detections. And \texttt{cross\_modal\_tension} finally flags windows where the modalities disagree.

This last signal is precisely the reason we perceive both modalities in one pass rather than run separate vision and audio detectors and merge their outputs. A modular pipeline sees each modality on its own and never compares them unless a component is added for that purpose, so a scene where calm imagery accompanies shouted audio, or where an officer's tone contradicts what the camera shows, produces two unremarkable detections and no flag at all.

\section{Summarization}
\label{app:summ}
After all windows of a video are perceived, we build a two-level narrative. Consecutive windows are first grouped into segments of ten, about five minutes of footage. For each segment the model then writes a short timestamp-anchored narrative from the per-window records, which is the map step. A final pass then composes a whole-video overview from the segment narratives and the category distribution, which is the reduce step. Every summarization prompt stays short and bounded, and because segment narratives are independent, we generate them as a batch in the multi-GPU cluster setting.

\section{Perception Engine}
\label{app:enginedetail}

\textbf{Speedup} \\
Let $\bar{\alpha}$ be the mean number of drafted tokens the model accepts per verification forward. Each forward then commits $\bar{\alpha} + 1$ tokens, against exactly one token for plain greedy decoding, so the ideal speedup is
\[
S \;=\; \frac{\bar{\alpha} + 1}{1 + c},
\]
where $c$ is the relative overhead of drafting and verification. Importantly, the scheme needs no draft model, no extra weights, and no additional memory.

\textbf{Batch controller} \\
With target batch $B^\ast = 8$ and current batch $B$, the controller applies
\[
B \;\leftarrow\;
\begin{cases}
\max\!\big(\lceil B/2 \rceil,\, 1\big) & \text{on OOM, retry},\\[2pt]
\min\!\big(B + 1,\, B^\ast\big) & \text{after 3 clean batches}.
\end{cases}
\]

\textbf{Chat preemption} \\
A background generation polls a flag. When a chat request is waiting for a replica and none is free, it then aborts the current window mid-decode, releases the replica, lets chat run, and afterwards redoes the window.\footnote{We do it this way because of the constraints we have. In an unconstrained setting, a more stable approach could be used to handle the chat.}

\textbf{Host memory} \\
In our case, the target partition caps resident RAM at 64 GB through a control group, and a naive thousand-video run is killed after a few dozen videos. The process memory footprint climbs for two reasons. First, the allocator retains freed frame and audio decode buffers in per-thread memory pools rather than returning them to the operating system, and second, page cache from reading source files is charged to the job's control group. Neither is released by standard cleanup.

We therefore added an explicit host-memory mechanism. After each video we return freed memory pools to the operating system, drop the just-read file from the page cache, and bound memory pool fragmentation. On top of this, the batch loop reads the job's actual control-group memory usage $M$ against the limit $M_{\max}$. Above a ceiling $\theta M_{\max}$ with $\theta = 0.82$, it then reclaims and then halves the batch for that round, and grows back when memory frees.

\section{Search Ranking}
\label{app:bm25}
For a query $q$ and a window document $d$,
\begin{align*}
\mathrm{score}(q,d) &= \sum_{t \in q} \mathrm{IDF}(t)\,
\frac{f(t,d)\,(k_1 + 1)}{f(t,d) + k_1\bigl(1 - b + b\,\frac{|d|}{\overline{|d|}}\bigr)},\\
\mathrm{IDF}(t) &= \ln\!\left( \frac{N - n(t) + 0.5}{n(t) + 0.5} + 1 \right),
\end{align*}
where $f(t,d)$ is the frequency of term $t$ in $d$, $|d|$ is the document length, $\overline{|d|}$ is the mean document length, $N$ is the number of indexed documents, $n(t)$ is the number that contain $t$, and $k_1 = 1.2$, $b = 0.75$.

\section{Audit Log}
\label{app:nonrep}
\begin{figure}[h]
\centering
\begin{tikzpicture}[
  node distance=3mm and 5mm,
  rec/.style={draw, rounded corners=2pt, align=center, minimum height=12mm, minimum width=20mm, font=\scriptsize, fill=blue!5},
  arr/.style={-{Stealth[length=2mm]}, thick}
]
\node[rec] (w0) {window $0$\\$o_0,\, t_0$\\$h_{-1}{=}g \to h_0$};
\node[rec, right=of w0] (w1) {window $1$\\$o_1,\, t_1$\\$h_0 \to h_1$};
\node[rec, right=of w1] (w2) {window $2$\\$o_2,\, t_2$\\$h_1 \to h_2$};
\node[right=of w2, font=\small] (dots) {$\cdots$};
\node[draw, rounded corners=2pt, above=of w1, font=\scriptsize, fill=orange!10, minimum width=24mm] (file) {source file hash $f$};
\draw[arr] (w0) -- (w1);
\draw[arr] (w1) -- (w2);
\draw[arr] (w2) -- (dots);
\draw[arr, dashed] (file) -- (w0);
\draw[arr, dashed] (file) -- (w1);
\draw[arr, dashed] (file) -- (w2);
\end{tikzpicture}
\caption{Per-video hash chain. Each window record hashes its raw model output, its timestamp, its index, the source file hash, and the previous hash.}
\label{fig:chain}
\end{figure}
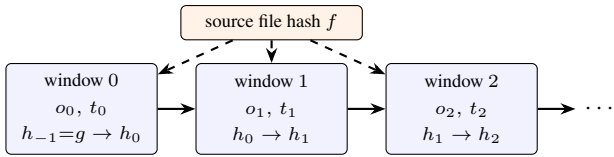

Entry $j$ of the global log stores
\begin{align*}
c_j &= \mathrm{SHA256}\big(\mathrm{canon}(\,\mathrm{actor}_j,\; \mathrm{action}_j,\; t_j\,)\big),\\
\ell_j &= \mathrm{SHA256}\big(\,c_j \,\|\, \ell_{j-1}\,\big),
\end{align*}
and each entry additionally records the acting officer's identity and a timestamp. The requirement of Keita and Homan \cite{keita2026nonrepudiation} is a tamper-evident record that binds a reported result to the computation that produced it, extended to every designated action of an autonomous agent, bound to its identity, its content, and its position in the session order.

Corrections in turn carry one of four actions (\emph{correct}, \emph{add}, \emph{reject}, \emph{accept}) together with the field changed and its old and new values. For hand-off, a single window then exports as a self-contained evidence packet holding the full record, an integrity block with the file hash, the chain position, and the verification status, a readable summary, and optionally the extracted clip.

Since $c_j$ covers the officer identity and the change, and because that hash is chained into every later entry through $\ell_j$, an officer cannot afterward claim that a correction was not theirs, that it said something different, or that it happened at a different point in the sequence, without breaking a chain that verification will localize. The same logic likewise protects the machine. The original perception output $o_i$ is hashed into the per-video chain, so the system cannot later be said to have produced a different finding than it did.

This matters for three reasons. First, accountability. An oversight body or a court can attribute every edit to a named reviewer and trust that the attribution cannot be forged or disowned. Second, defense of the record. A party cannot allege after the fact that findings were fabricated or quietly rewritten, because any such change is detectable and localizable. Third, chain of custody. The combination of source file hash, per-window content hashes, and an attributed audit trail together gives a continuous, verifiable custody record from ingest through every human touch.

\section{Window Record}
\label{app:schema}
We store each window as a row whose principal columns, with JSON-typed fields marked $\dagger$, are \texttt{chunk\_id}, \texttt{source\_video}, \texttt{source\_path}, \texttt{file\_hash}, \texttt{timestamp\_start}, \texttt{timestamp\_end}, \texttt{chunk\_index}, \texttt{duration\_sec}, \texttt{description}, \texttt{audio\_description}, \texttt{cross\_modal}, \texttt{transcript}, \texttt{speakers}$\dagger$, \texttt{cluster}, \texttt{cluster\_scores}$\dagger$, \texttt{ai\_confirmed}, \texttt{confidence}, \texttt{ambiguity\_note}, \texttt{tags}$\dagger$, \texttt{audio\_events}$\dagger$, \texttt{evidence\_flags}$\dagger$, \texttt{persons\_count}, \texttt{has\_weapon}, \texttt{weapon\_drawn}, \texttt{physical\_contact}, \texttt{anomaly\_score}, \texttt{cross\_modal\_tension}, \texttt{environment}, \texttt{camera\_type}, \texttt{officer\_id}, \texttt{parent\_segment}, \texttt{parent\_video}, \texttt{has\_audio}, \texttt{has\_transcript}, \texttt{prev\_doc\_hash}, \texttt{this\_doc\_hash}, \texttt{ingested\_at}, \texttt{ingest\_week}, \texttt{shift\_date}, \texttt{engine\_variant}. Note that the columns \texttt{chunk\_id}, \texttt{chunk\_index}, \texttt{cluster}, and \texttt{cluster\_scores} carry historical names: \texttt{chunk} denotes what the text calls a window, and \texttt{cluster} holds the classification label of Section~\ref{sec:pipeline} rather than the output of any clustering procedure.

Table~\ref{tab:schema} summarizes the key fields. Videos and segments in turn carry analogous aggregate rows, namely dominant category, category distribution, peak-anomaly window, critical-window lists, overview narrative, and a training-value label.

\begin{table}[h]
\centering
\small
\begin{tabular}{@{}ll@{}}
\toprule
Field & Meaning \\
\midrule
\texttt{description} & visual narrative with offsets \\
\texttt{audio\_description} & what is heard \\
\texttt{cross\_modal} & vision/audio agreement \\
\texttt{transcript} & verbatim, speaker-tagged \\
\texttt{cluster} & triage category (3 labels) \\
\texttt{cluster\_scores} & 3 floats summing to 1 \\
\texttt{confidence} & high, medium, or low \\
\texttt{anomaly\_score} & unusualness in $[0,1]$ \\
\texttt{evidence\_flags} & e.g.\ \texttt{weapon\_drawn} \\
\texttt{has\_weapon} & structured boolean \\
\texttt{physical\_contact} & officer-civilian contact \\
\texttt{cross\_modal\_tension} & contradictory affect \\
\texttt{prev\_doc\_hash} & chain link (Sec.~\ref{sec:integrity}) \\
\texttt{this\_doc\_hash} & chain link (Sec.~\ref{sec:integrity}) \\
\bottomrule
\end{tabular}
\caption{The per-window record, selected fields.}
\label{tab:schema}
\end{table}

\section{Settings}
\label{app:hyper}

\begin{table}[h]
\centering
\small
\begin{tabular}{@{}ll@{}}
\toprule
Parameter & Default \\
\midrule
Window, overlap, stride & 30s, 5s, 25s \\
Frame sampling rate & 1 fps (30 per window) \\
Audio & 16 kHz mono \\
Windows per segment & 10 ($\sim$5min) \\
Perception output cap & 2048 tokens \\
Summary output cap & 768 tokens \\
Anomaly spike threshold & 0.70 \\
Speculative $K_s$, $n$-gram & 12, 2 \\
BM25 $k_1$, $b$ & 1.2, 0.75 \\
Attention kernel & SDPA \\
Consumer precision & 4-bit NF4 (QAT) \\
Cluster precision & bf16 \\
Cluster batch target $B^\ast$ & 8 (AIMD) \\
Host-memory ceiling $\theta$ & 0.82 of cgroup limit \\
Data parallelism & one replica per GPU \\
Agent steps, re-perceptions & 8, 6 per question \\
Context length & 256K tokens \\
\bottomrule
\end{tabular}
\caption{Default settings.}
\label{tab:hyper}
\end{table}

\section{Constraints}
\label{app:eff}

\begin{table}[h]
\centering
\small
\begin{tabular}{@{}p{0.30\linewidth}p{0.60\linewidth}@{}}
\toprule
Binding constraint & Mechanism \\
\midrule
16 GB GPU (capacity) & 4-bit NF4 QAT weights, embedders and head in higher precision \\
Per-GPU decode speed & Lossless temporal speculative decoding, about $1.48\times$ \\
Multi-GPU scaling & One data-parallel replica per GPU, about $N\times$ \\
H100 & bf16 with per-window batching \\
Multiprogrammed GPUs & AIMD batch controller \\
64 GB host cap & Memory pool trim, page-cache drop, cgroup-aware backoff \\
Preemption and crashes & Incremental commits, resumable ingestion \\
Chat during ingest & Window-granular preemption \\
\bottomrule
\end{tabular}
\caption{Constraints and the mechanisms for them.}
\label{tab:eff}
\end{table}

\section{Speculative Decoding}
\label{app:spec}
We give below the decoding loop we run per replica. Let $y^{(i-1)}$ be the full token sequence emitted for the previous window in this replica's stream.

\begin{algorithm}[h]
\caption{Lossless temporal speculative decoding for window $i$}
\label{alg:spec}
\textbf{Input}: window frames, audio, prompt; reference $y^{(i-1)}$\\
\textbf{Output}: token sequence $y$, byte-identical to greedy decoding
\begin{algorithmic}[1]
\STATE $y \gets$ empty; feed the window's frames, audio, and prompt
\WHILE{not end-of-sequence and $|y| <$ output cap}
  \STATE find the last $n{=}2$ tokens of $y$ in $y^{(i-1)}$
  \IF{a match exists}
    \STATE draft $d \gets$ up to $K_s{=}12$ tokens following the match
  \ELSE
    \STATE draft $d \gets$ empty
  \ENDIF
  \STATE run one forward over $y \,\|\, d$; get greedy predictions
  \STATE accept the longest prefix of $d$ matching the predictions
  \STATE append the accepted prefix and the next token to $y$
\ENDWHILE
\STATE \textbf{return} $y$
\end{algorithmic}
\end{algorithm}

Acceptance is high, as consecutive windows overlap by 5 seconds and describe the same scene, officer, environment, and JSON scaffolding. The JSON keys, which every record repeats, are in particular accepted almost for free. The first window of each stream, by contrast, runs plain greedy decoding, since it has no reference.

\section{Detectors}
\label{app:detectors}
\textbf{Agent path test.} On the two recordings of Section~\ref{sec:experiment} we ran an eight-question supervisory review form used by the department.\footnote{The questions are kept private following the guidelines of the department.} All sixteen answers carried window-level citations with timestamps, so no claim reached the reviewer without a window the model had re-perceived in that turn. One experienced reviewer, reading the outputs, then asked \textbf{``whether the system had been trained on police use-of-force reports.''}

To bridge the gap between how a user phrases an event and how the model tagged it, we further maintain a catalog of \emph{detectors} such as taser, firearm-drawn, gunshot, use-of-force, Miranda warning, K9, and foot pursuit. A find-event query then runs a three-stage candidate search and afterwards confirms each candidate by re-perceiving it in the detector's modality, returning per-window yes, no, or unsure verdicts. Confirmed windows, and only those, are citable. Appendix~\ref{app:detectors} lists the catalog and gives a further worked example.

\textbf{A worked query.} Suppose a reviewer asks whether any officer drew a firearm during a given shift. The agent first issues one SQL statement combining the structured prefilter \texttt{weapon\_drawn = true} with BM25 terms such as \emph{gun}, \emph{holster}, and \emph{pointed}, which returns, say, nine candidate windows. It then re-perceives the six highest-ranked of these, passing each window's frames and audio back to the model with a prompt asking specifically whether a firearm is drawn and by whom; four come back \emph{yes}, one \emph{no}, and one \emph{unsure}. The agent next drafts an answer citing five windows, but one of those was surfaced by search and never re-perceived, so the grounding filter removes that marker and the sentence resting on it. As a result, the reviewer receives a four-citation answer, each citation opening the exact 30 seconds of footage it rests on, and the audit log gains one entry recording the question, the six re-perceived window identifiers, and content hashes over the re-perception outputs and the final answer.

Each detector maps a family of aliases to high-precision structured prefilters, high-recall lexical terms, and a preferred modality with a focused re-perception prompt. The catalog includes taser and CEW, firearm-drawn, gunshot, siren, emergency-lights, handcuffing, Miranda warning, use-of-force, foot-pursuit, verbal-commands, K9, medical-aid, vehicle-collision, screaming and distress, de-escalation, and door-breach, among others. An unrecognized query, meanwhile, falls back to a generic both-modality detector built from the user's own terms. A find-event call generates candidates in three stages (structured flags, then synonym text, then a top-anomaly fallback), and then confirms each by re-perception before it is reported or cited.

As one worked example, the firearm-drawn detector prefilters on \texttt{weapon\_drawn = true} or the \texttt{weapon\_drawn} evidence flag. It then expands lexically to terms such as \emph{gun}, \emph{pistol}, \emph{firearm}, \emph{drew}, \emph{holster}, and \emph{pointed}, and it prefers the visual modality. It finally re-perceives each candidate with a prompt that asks whether a firearm is drawn and by whom, returning yes, no, or unsure with a one-sentence justification.

\section{Transcription}
\label{app:transcribe}
Standalone diarized transcription re-listens to each window and produces speaker-labeled, timestamped, confidence-tagged entries. We also carry a running speaker registry across a video. Labels and audio-and-visual speaker descriptions created in earlier windows are supplied to later windows, so a voice keeps a stable identity (\texttt{OFFICER\_1}, \texttt{DISPATCH}, and so on) across the recording, and the model in turn grounds who is speaking in the visible scene. An optional verification pass then re-examines each drafted window against the original audio under explicit faithfulness rules, namely no invented dialogue, \texttt{[INAUDIBLE]} for anything not clearly audible, and integrity of spoken identifiers such as addresses and call signs. Transcripts can finally be exported to SRT, TXT, and JSON.

\section{Reports}
\label{app:analysis}
Over the window store, we compute several review-oriented views without further perception. The incident timeline lists significant windows in order. The safety summary in turn groups windows by officer-safety category (force, weapon, restraint, pursuit, medical, and so on). The policy check then pairs each indicated safety category with the governing sections of ingested policy documents and the compliance question a reviewer must answer. The redaction worksheet finally flags windows likely to contain personally identifiable information (faces, minors, spoken names, plates, addresses, and phone numbers) for public-records release, and it over-flags on purpose.

Finally, we generate debrief reports by compiling a scope's critical windows with any officer corrections applied, then prompting the model for a sectioned narrative that cites windows. We then commit the report with a content hash and an audit entry so it is reproducible.

\section{Transcription, Analysis, and Operations}
\label{app:ops}
OmniEye is not yet in daily production use. We are instead rolling it out into the Rochester Police Department in stages: processing the archived recordings on department infrastructure with reviewers annotating the output, then training key staff to use the tool to identify training gaps, then wiring it into the department's existing analysis workflow, and finally making it available to other law-enforcement agencies.

Standalone diarized transcription re-listens to each window and produces speaker-labeled, timestamped, confidence-tagged entries, with a running speaker registry that keeps a voice's identity stable across a recording (Appendix~\ref{app:transcribe}). Over the same window store we then compute review-oriented views without further perception: an incident timeline, a safety summary, a policy check against ingested documents, a redaction worksheet, and reproducible debrief reports (Appendix~\ref{app:analysis}).

Ingestion also tolerates per-file failure, so a bad or oversized file is logged and skipped rather than blocking the run; in the 1,000-recording experiment of Section~\ref{sec:experiment}, for instance, seven zero-length files were skipped this way without interrupting ingestion. It moreover commits incrementally, and it records unfinished jobs in a recovery registry keyed to the workspace. After a crash or a preemption, the same command therefore re-attaches to the workspace and continues. Since window identifiers are deterministic and already-written windows are skipped, resumption redoes only what is missing. Jobs can likewise be paused and resumed, and a paused or interrupted job survives being saved into a portable bundle and continued on another machine.

The system is also fully usable without a display. Single commands ingest a folder, list and resume unfinished work, export the recordings catalog as JSON, and select the compute precision (4-bit or bf16) for a run. Beyond that, we expose the same capabilities in a mouse-first, keyboard-accelerated terminal application designed for non-technical reviewers. It has a conversation panel and a tool trace, a jobs panel, a file browser, a metrics dashboard, and an annotation and review screen with clickable, playable citations. For performance, we finally use a \textbf{Rust data layer} \cite{pyo3} that accelerates scanning, hashing, media extraction, and bundle packing, with pure-Python fallbacks so the tool runs before the native extension is built.

\section{Commands}
\label{app:cli}
The headless interface exposes the full workflow in single commands. These commands ingest a folder into a workspace, list unfinished jobs, resume a job by identifier, and pause a running job. They also export the recordings catalog as JSON, export a workspace as a full or database-only \texttt{.omnieye} bundle, and load a bundle. They further verify a video's hash chain, verify the audit log, and export an evidence packet for a window. They finally run a stratified sample selection with a fixed seed, and select the compute precision (4-bit or bf16) for a run. Every command is safe to re-run. Ingestion skips completed windows, bundle export is deterministic given the database state, and verification is read-only.

\section{Verification}
\label{app:integrity}
Per-video chain verification passes a video's windows in index order from the genesis hash, and asserts, for each window $i$, that its stored predecessor hash equals the actual predecessor's content hash,
\[
\forall i \ge 1, \quad \mathrm{stored\_prev}(i) \stackrel{?}{=} h_{i-1},
\]
with $\mathrm{stored\_prev}(0) \stackrel{?}{=} g$. Any mismatch therefore localizes a broken link to a window index. Audit verification in turn walks the global log in sequence, recomputing both the inter-entry link $\ell_j$ and each entry's content hash $c_j$, and it then reports broken linkage or content mismatch by sequence number. Each audit entry's content hash covers the acting officer's identity, and, for agent actions, the agent's re-perceived window set $R_t$ and answer hash (Section~\ref{sec:agent}). A passing verification therefore also certifies that every recorded action is correctly attributed and cannot be repudiated by either a human or the agent. Three denials in particular are ruled out for the agent. First, the agent, or an operator on its behalf, cannot later deny that a given answer was produced. Second, no party can claim the agent cited a window it never examined, because a citation in $y_t'$ means the window is in the logged set $R_t$, and $R_t$ is hash-bound. And third, no party can claim the agent saw different pixels than it did, because the cited window's record sits in a per-video chain rooted in the source file hash $f$. Both procedures return a boolean and a list of localized issues, suitable for display in an evidence packet.

\bibliography{custom1}

\end{document}